\documentclass[epj,nopacs]{svjour}
\usepackage{epsfig,array}
\usepackage{multirow,color}
\usepackage{graphicx}
\newcommand{\be}{\begin{eqnarray}}
\newcommand{\ee}{\end{eqnarray}}
\newcommand{\oo}{$^o$}
\newcommand{\er}{$\pm$}

\newcommand{\bea}{\begin{eqnarray}}
\newcommand{\eea}{\end{eqnarray}}

\newcommand{\beq}{\begin{equation}}
\newcommand{\eeq}{\end{equation}}

\def\fun#1#2{\lower3.6pt\vbox{\baselineskip0pt\lineskip.9pt
\ialign{$\mathsurround=0pt#1\hfil##\hfil$\crcr#2\crcr\sim\crcr}}}

\begin{document}

\title{\boldmath Helicity amplitudes for photoexcitation of nucleon resonances off neutrons}
\titlerunning{Helicity amplitudes for photoexcitation of nucleon resonances off neutrons}
\author{A.V.~Anisovich$\,^{1,2}$,  V. Burkert$\,^3$, E.~Klempt$\,^1$,
  V.A.~Nikonov$\,^{1,2}$,  A.V.~Sarantsev$\,^{1,2}$, U.~Thoma$\,^1$
}
\authorrunning{A.V.~Anisovich \it et al.}
\institute{$^1\,$Helmholtz-Institut f\"ur Strahlen- und Kernphysik,
Universit\"at Bonn, Germany\\
$^2\,$Petersburg Nuclear Physics Institute, Gatchina, Russia\\
$^3\,$Jefferson Lab, 12000 Jefferson Avenue, Newport News, Virginia,
US }
\date{Received: \today / Revised version:}
\mail{klempt@hiskp.uni-bonn.de} \abstract{The helicity amplitudes
$A^{1/2}_n$ and $A^{3/2}_n$ for the photoexcitation of nucleon
resonances off neutrons are determined in a multi-channel
partial-wave analysis.
 \vspace{1mm}   \\
} \maketitle

\section{Introduction}

One of the most intriguing issues in the physics of strong
interactions is the spectrum  and internal structure of baryon
resonances \cite{Hey:1982aj,Klempt:2009pi,Aznauryan:2012}. In the
calculation of mesonic excitations, quark models have been very
successful \cite{Klempt:2007cp,Anisovich:2008zz}. Classical examples
are the non-relativistic quark model \cite{Gunion:1975yx} and its
relativized \cite{Godfrey:1985xj} or fully relativistic versions
\cite{Koll:2000ke,Ricken:2000kf}. These models use a linear
confinement potential simulating confinement and some residual
quark-antiquark interactions. But in the baryon sector, the
situation is much more complicated. In the low mass region, the
interpretation of baryons as three-quark bound states in a linear
confinement potential appears to be rather successful, even  though
important details such as the masses of radial excitations like the
Roper resonance $N(1440)1/2^+$ remain unexplained
\cite{Capstick:1986bm,Stassart:1997vk,Loring:2001kx}. The choice of
a different form of the confinement potential \cite{Ferraris:1995ui}
or of the residual interaction \cite{Glozman:1997ag} improves the
mass pattern. More recently, dynamical approaches such as the
dynamically coupled channel models \cite{Suzuki:2010}, that include
coupling to inelastic channels, have been successful in predicting
the correct pole structure of this state. At higher masses quark
models predict a large number of states which were not found
experimentally up to now. This problem triggered two developments:
the start of new experiments searching for these {\it missing
resonances}, and elaboration of new theoretical models which can
explain the observed spectrum.

The most direct approach to calculate the masses of baryon
excitations is to discretize QCD on a lattice
\cite{Edwards:2011jj,Edwards:2012fx}. It turns out, however, that
the number of states is identical to the number expected in a simple
picture where the three-body system is described by two harmonic
oscillators, at least in the first and second excitation band. In
the low-energy region, the mass pattern resembles the observed
spectrum. At the present stage, lattice QCD computations still use
unphysical quark masses, and do not include decays of excited
states. They therefore provide no explanation why a large fraction
of the states remains unobserved. Also, the predicted mass spectrum
does not agree with observations.

Possibly, quarks are not the only degrees of freedom to describe
baryon resonances, they may possibly be better described by the
interaction of stable (or quasi-stable) hadrons, from the
interaction of  mesons and baryons. The interaction is then derived
from chiral Lagrangians, from an effective theory of QCD at low
energies
\cite{Weinberg:1978kz,Meissner,Lutz,Bruns:2010sv,Mai:2012wy}. Great
progress has been made in a better understanding of the properties
of a few baryon resonances. At present, it is however not yet clear
what spectrum emerges in this approach, and it is even unclear if
these dynamically generated states come atop of quark model states
or if they reflect a better understanding of quark model states.
Measurements of helicity amplitudes will be essential in trying to
discern the internal structure of these states.

Up to recently, our experimental knowledge of the baryon spectrum
was based mostly on the classical analyses
\cite{Hohler:1979yr,Cutkosky:1980rh} at Karlsruhe and
Carnegie-Mellon and on the more recent analysis at GWU
\cite{Arndt:2006bf} of data on elastic $\pi N$ scattering. The large
inelasticity in all partial waves demonstrates the importance of
inelastic reactions. Yet, pion-induced inelastic data have rather
poor statistic and, sometimes, there are contradictions between
different experiments. But there is a wealth on new data from
photoproduction experiments \cite{Burkert:2012ee}. A multichannel
partial wave analysis of these data allowed us to confirm the
existence of resonances which were ambiguously identified in the
analysis of the elastic $\pi N$ data, to observe new states, and to
define many properties of baryons with a good precision
\cite{Anisovich:2011fc}. The analysis hence yields important
information on resonance couplings to different final states but
also on the $\gamma N$ channel. These couplings depend on the
internal structure of baryons and can be calculated in the framework
of different models. A list of the $\gamma p$ and $\gamma n$
couplings calculated in the framework of a relativized quark model
can be found in \cite{Capstick:2000qj}. The agreement between
calculated and observed values is one of the remarkable successes of
the quark model. Other models do not yet provide the full list of
the $\gamma N$ couplings although for separate states such
calculations can be found in the literature: see, for example, the
calculation of the Roper $\gamma N$ couplings with the AdS/QCD
approach\cite{Gutsche:2012wb}. There is no doubt that a relevant
model should reproduce not only masses of baryons but also their
decay properties.

Knowledge of the resonance photocouplings can give insight into the
underlying symmetry and internal structure of the states. For
example, the Single Quark Transition (SQT) model
\cite{Hey:1974,Cottingham:1979} describes the electromagnetic
transitions from the ground state nucleon to all states within the
$SU(6)$ multiplets $(70,1^-)_1$ and $(56,2^+)_2$ with just 3 and 4
reduced amplitudes, respectively. With accurate data on the
photocoupling amplitudes for these states, these relationships can
be tested and provide further insights into the underlying symmetry.
Surprisingly good results were found for the $(70,1^-)_1$ multiplet
\cite{Burkert:2003}, for which the reduced amplitudes were
determined from the well-measured $N(1535)1/2^-$ and $N(1520)3/2^-$
photocoupling amplitudes on protons, and used to predict the
remaining 14 amplitudes for states in this multiplet on both protons
and neutrons. Strong deviations from these predictions for
individual states would indicate that contributions other than
single quark transitions may contribute or that the assignment of
the state to a specific $(D, L_{\tt N}^P)$ multiplet is
questionable.

In this paper we report the result of the combined analysis of the
data base used for solutions BG2011 \cite{Anisovich:2011fc} with
data on $\pi^-$, $\pi^0$, and $\eta$ photoproduction off neutron.
The data are not sufficiently precise to increase our knowledge of
masses, widths and hadronic decay modes of baryon resonances. Hence
all these quantities can remain frozen, and only the $\gamma n$
helicity couplings are free parameters of the fit. In a final fit
all parameters are set free; this gives a slight improvement in
$\chi^2$ without significant changes in any of the parameters. The
fit returns these helicity amplitudes. They are tabulated here and
compared with previous determinations.

\section{Data base}

The data used for the determination of the $\gamma n$ couplings of
the $N^*$ states are listed in Table~\ref{gamma_n_data}. In addition
to the new JLab data on $\gamma n\to p\pi^-$
\cite{Chen:2009sda,Chen:2012yv} a
large number of further data on this reaction \cite{gn_pion_md_dcs1,%
gn_pion_md_dcs2,gn_pion_md_dcs3,gn_pion_md_dcs4,gn_pion_md_dcs5,%
gn_pion_md_dcs6} has been included in the fit.

In bubble chambers all outgoing charged particles are detected
\cite{gn_pion_mp_dcs0_benz,gn_pion_mp_dcs0_Rossi} and the events are
kinematically fully reconstructed.  No correction for Fermi motion
is hence required. We give the $\chi^2$ contribution for the latter
data separately. Several groups have reported measurements on the
beam asymmetry $\Sigma$ using linearly polarized
photons \cite{gn_pion_md_sigm,gn_pion_md_sig2,gn_pion_md_sig3,%
gn_pion_md_sig4,gn_pion_md_sig5,gn_pion_md_sig6,%
gn_pion_md_sig7,gn_pion_md_sig8,gn_pion_md_sig9} and the target
asymmetry $T$ using a transversely polarized deuteron target
\cite{gn_pion_md_tpol,gn_pion_md_tpo2,gn_pion_md_tpo3,gn_pion_md_tpo4}.
The recoil polarization of the proton $P$ was determined by
exploiting the analyzing power of the proton-Carbon scattering
process \cite{gn_pion_md_ppol,gn_pion_md_ppo2}.

The reverse reaction $\pi^-p\to n\gamma$ has the advantage that no
Fermi correction needs to be applied. Data were taken at several
laboratories \cite{gn_pion_mp_dcs0,gn_pion_mp_dcs1,gn_pion_mp_dcs2,%
gn_pion_mp_dcs3,gn_pion_mp_dcs4,gn_pion_mp_dcs5,gn_pion_mp_dcs6}. By
polarizing the target protons, the polarization variable $P$ is
determined \cite{gn_pion_mp_ppol,gn_pion_mp_ppo1}.

Differential cross sections for photoproduction of neutral pions off
neutrons $\gamma d\to n\pi^0 (p)$ are obtained using a deuteron
target \cite{gn_pion_pd_said,gn_pion_pd_2,gn_pion_pd_3,%
gn_pion_pd_4}. The beam asymmetry $\Sigma$ for this reaction is
available from \cite{gn_pion_pd_gra1}. Similar data exist for
$\gamma d\to \eta\gamma (p)$ \cite{gn_eta0_pd_els2,gn_eta0_pd_gra1}.

Data obtained with a deuteron target are affected by the Fermi
motion of the neutron and by rescattering in the final state (Final
State Interaction, FSI). The results thus depend on the cuts made in
the experiment or the analysis: for example, a particular cut on the
momentum of the spectator proton influences the smearing of the data
due to Fermi motion and the contribution from the FSI.

\begin{table}[pt]
\caption{\label{gamma_n_data}Data base for meson production off
neutrons as compiled by Data Analysis Center at GWU \cite{GWU}. The
table lists the reaction and reference to the original work, the
observable, the number of data points, and two $\chi^2$ values.
$\chi^2_0$ shows the quality of the fit using solution BG2011-02
with $\gamma n$ couplings as free parameters only. The $\chi^2_f$
shows the quality of the fit to the $\gamma n$ data after
optimization of all parameters. }
\renewcommand{\arraystretch}{1.2}
\begin{center}
\begin{tabular}{ccccc}
\hline \hline
$\gamma n\to\pi^-p$&Observ.&$N_{\rm data}$&$\chi_{0}^2$&$\chi^2_f$\\
\hline
\cite{Chen:2012yv,gn_pion_md_dcs1,gn_pion_md_dcs2,gn_pion_md_dcs3,gn_pion_md_dcs4,%
gn_pion_md_dcs5,gn_pion_md_dcs6}
&$d\sigma/d\Omega$&1298 & 2.84 & 2.32 \\
\cite{gn_pion_mp_dcs0_benz,gn_pion_mp_dcs0_Rossi}&$d\sigma/d\Omega$& 529 & 3.16 & 3.08 \\
\cite{gn_pion_md_sigm,gn_pion_md_sig2,gn_pion_md_sig3,%
gn_pion_md_sig4,gn_pion_md_sig5,gn_pion_md_sig6,%
gn_pion_md_sig7,gn_pion_md_sig8,gn_pion_md_sig9}
&$\Sigma$         & 316 & 3.74 & 3.08 \\
\cite{gn_pion_md_tpol,gn_pion_md_tpo2,gn_pion_md_tpo3,gn_pion_md_tpo4}
&$T$              & 105 & 4.96 & 3.18 \\
\cite{gn_pion_md_ppol,gn_pion_md_ppo2}&$P$     &  20 & 3.22 & 3.17 \\
\hline
$\pi^-p\to \gamma n$  &&&&\\
\hline
\cite{gn_pion_mp_dcs0,gn_pion_mp_dcs1,gn_pion_mp_dcs2,%
gn_pion_mp_dcs3,gn_pion_mp_dcs4,gn_pion_mp_dcs5,gn_pion_mp_dcs6}
&$d\sigma/d\Omega$& 495 & 1.65 & 1.53 \\
\cite{gn_pion_mp_ppol,gn_pion_mp_ppo1}&$P$&  55 & 4.59 & 3.11 \\
\hline
$\gamma d\to\pi^0 n(p)$&&&&\\
\hline \cite{gn_pion_pd_said,gn_pion_pd_2,gn_pion_pd_3,gn_pion_pd_4}
&$d\sigma/d\Omega$& 147 & 3.14 & 2.98 \\
\cite{gn_pion_pd_gra1}&$\Sigma$         & 216 & 2.82 & 1.90 \\
\hline
$\gamma d\to\eta n(p)$&&&& \\
\hline
\cite{gn_eta0_pd_els2}&$d\sigma/d\Omega$& 330 &1.57  & 1.40 \\
\cite{gn_eta0_pd_gra1}&$\Sigma$         &  88 & 2.42 & 2.17 \\
\hline \hline
\end{tabular}
\end{center}
\end{table}
The treatment of the Fermi motion in our approach was described in
detail in \cite{Anisovich:2008wd}. For the FSI we use the
corrections calculated by the SAID group
\cite{Chen:2012yv,Tarasov:2011ec}. Here, we have restricted the
fitted energy interval to invariant masses below 2.3\,GeV since
above, too little is known about photoproduction of resonances. The
data on $\eta$ photoproduction data off neutrons seem to suffer less
from FSI. The $\gamma d\to \eta p (n)$ cross sections agree very
well with data on $\gamma p\to \eta p$. Thus here we follow the
prescription of paper \cite{gn_eta0_pd_els2} where the difference
between two approaches to allow for FSI was included as a systematic
error.


\begin{figure*}
\begin{center}
\begin{tabular}{cc}
\epsfig{file=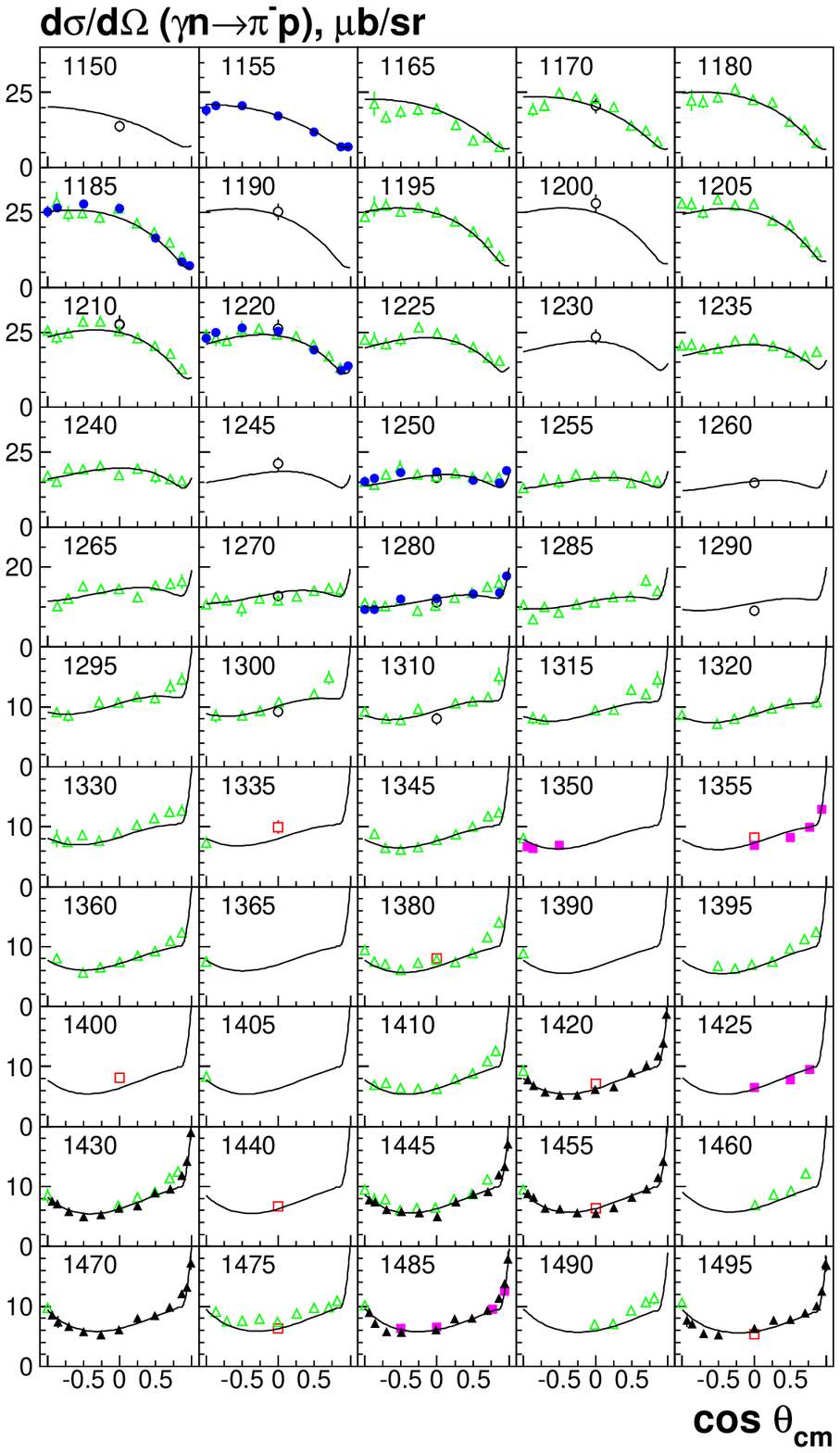,width=0.48\textwidth,height=0.58\textheight}&
\epsfig{file=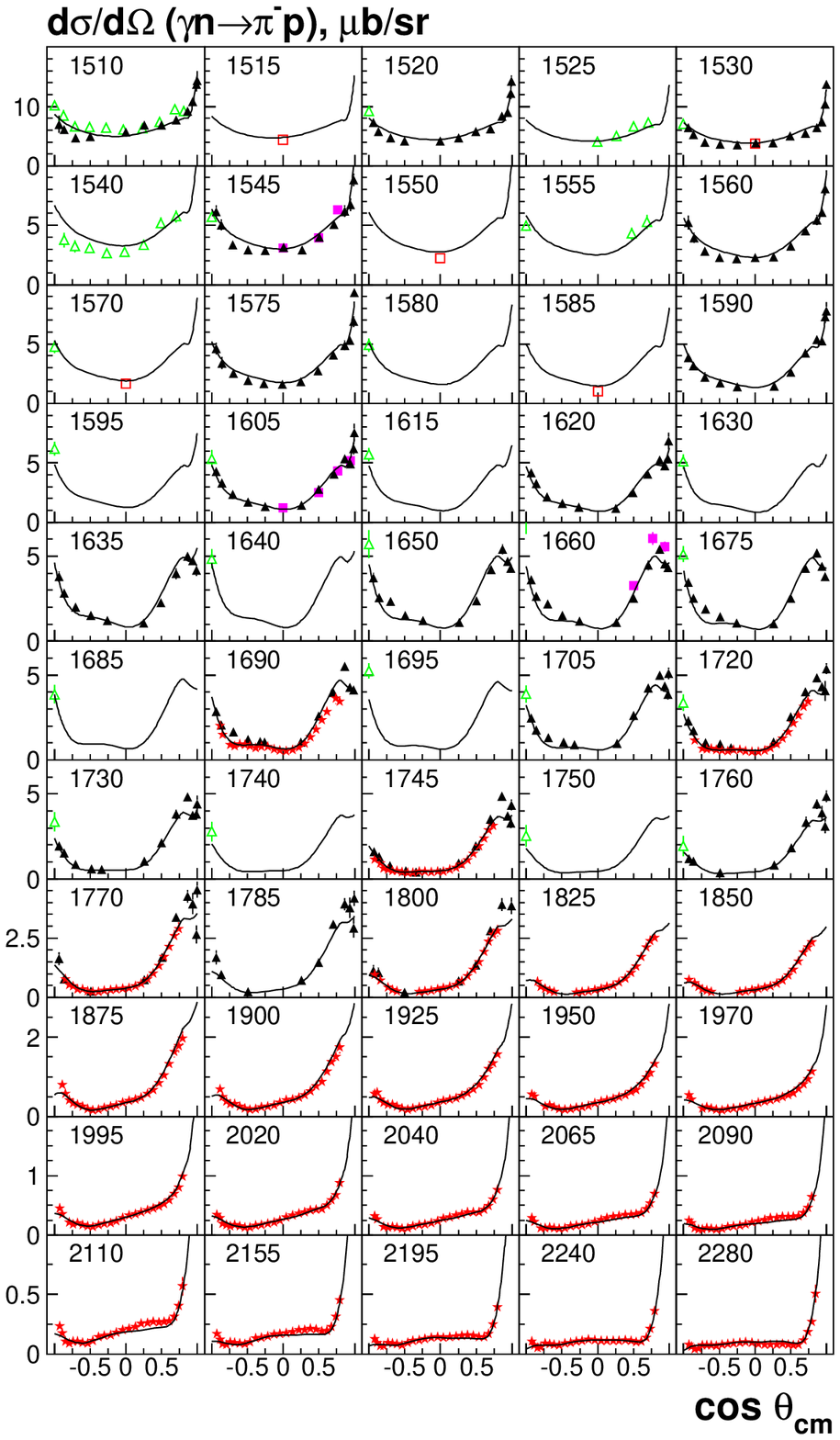,width=0.48\textwidth,height=0.58\textheight}
\end{tabular}
\end{center}
\caption{\label{gn_pimp_fm_dcs_1}Differential cross section for
$\gamma n\to\pi^- p$. Data:  red   filled stars   are from
\cite{Chen:2012yv}, black open   circles are from
\cite{gn_pion_md_dcs1}, red   open squares are from
\cite{gn_pion_md_dcs2}, green open triangles are from
\cite{gn_pion_md_dcs3}, blue  filled circles are from
\cite{gn_pion_md_dcs4}, pink  filled squares are from
\cite{gn_pion_md_dcs5}, black filled triangles are from
\cite{gn_pion_md_dcs6}. }
\begin{center}
\begin{tabular}{cc}
\epsfig{file=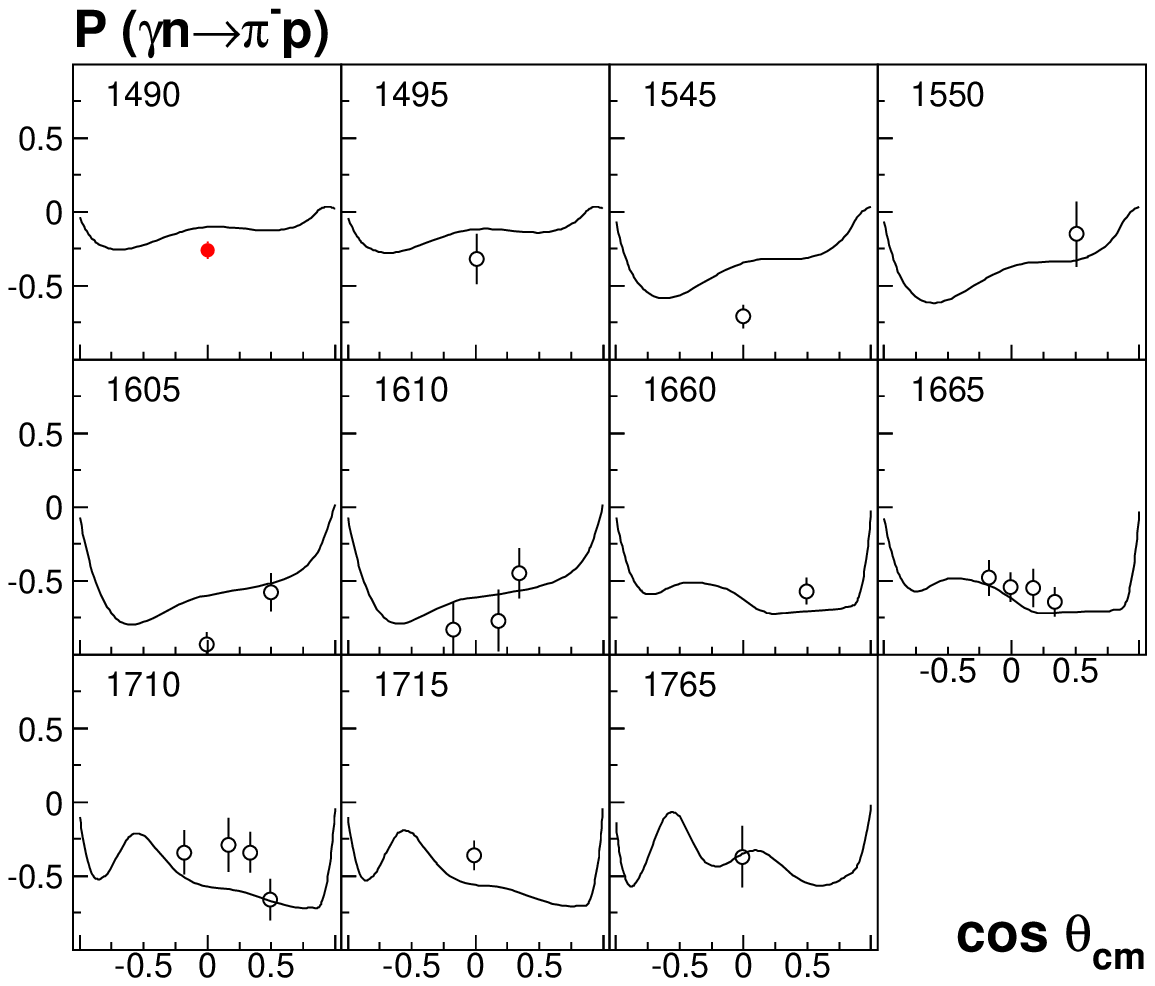,width=0.48\textwidth,height=0.25\textheight}&
\epsfig{file=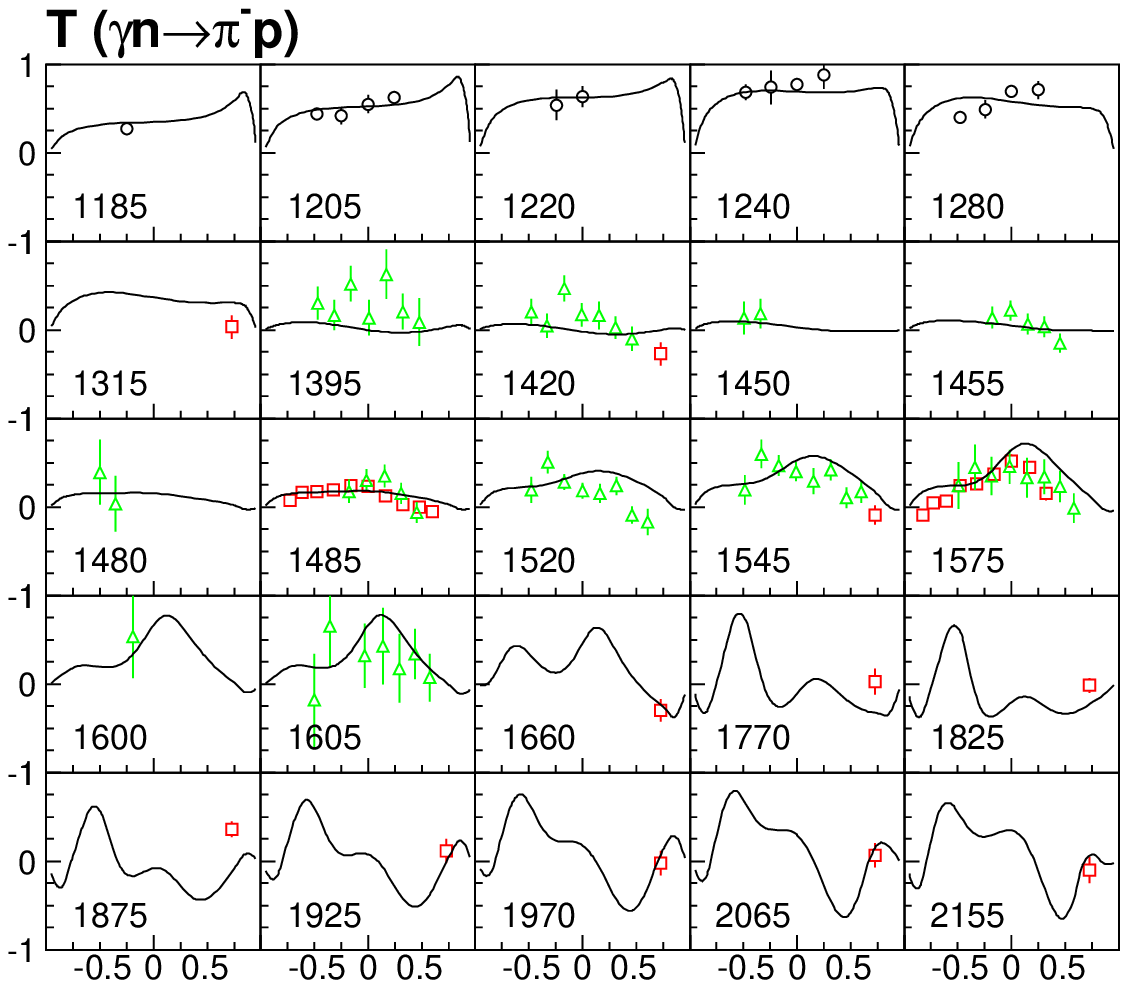,width=0.48\textwidth,height=0.25\textheight}
\end{tabular}
\end{center}
\caption{\label{gn_pimp_fm_ppol}Left: Recoil polarization $P$ for
$\gamma n\to\pi^- p$. Data: black open   circles are from
\cite{gn_pion_md_ppol}, red filled circles are from
\cite{gn_pion_md_ppo2}. Right: Target polarization $T$ of $\gamma
n\to\pi^- p$ reaction. Data: black open   circles are from
\cite{gn_pion_md_tpol}, red open squares are from
\cite{gn_pion_md_tpo2,gn_pion_md_tpo3}. Green open triangles are
from \cite{gn_pion_md_tpo4}.}
\end{figure*}

\begin{figure*}
\begin{center}
\begin{tabular}{cc}
\epsfig{file=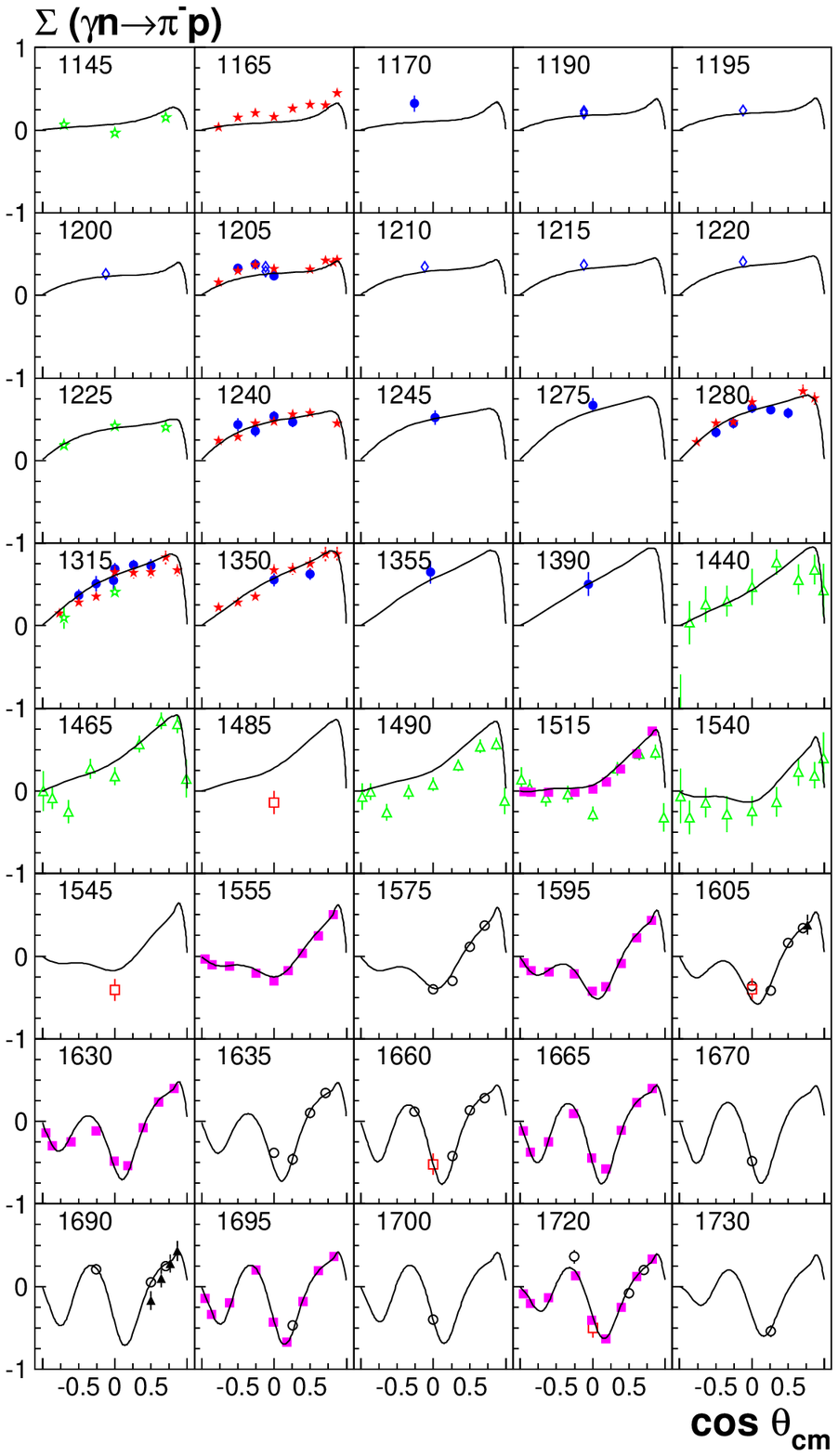,width=0.48\textwidth,height=0.57\textheight}&
\epsfig{file=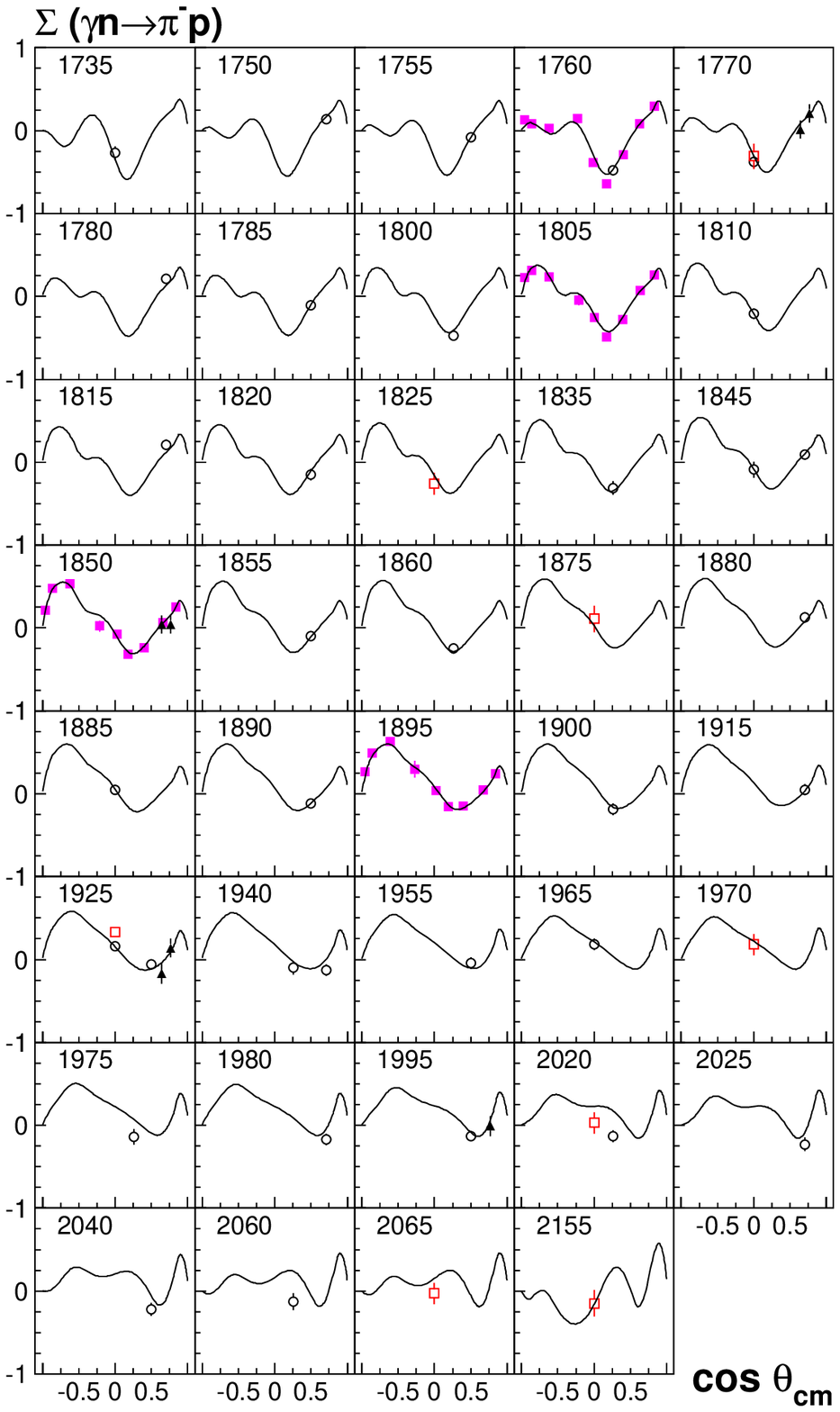,width=0.48\textwidth,height=0.57\textheight}
\end{tabular}
\end{center}
\caption{\label{gn_pimp_fm_sigm_1}$\Sigma$ polarization for $\gamma
n\to\pi^- p$. Data: black open   circles are from
\cite{gn_pion_md_sigm}, red open squares are from
\cite{gn_pion_md_sig2}, green open triangles are from
\cite{gn_pion_md_sig3}, blue  filled circles are from
\cite{gn_pion_md_sig4}, pink  filled squares are from
\cite{gn_pion_md_sig5}, black filled triangles are from
\cite{gn_pion_md_sig6}, red   filled stars   are from
\cite{gn_pion_md_sig7}, green open   stars   are from
\cite{gn_pion_md_sig8}, blue  open diamonds  are from
\cite{gn_pion_md_sig9}. }
\begin{center}
\begin{tabular}{cc}
\epsfig{file=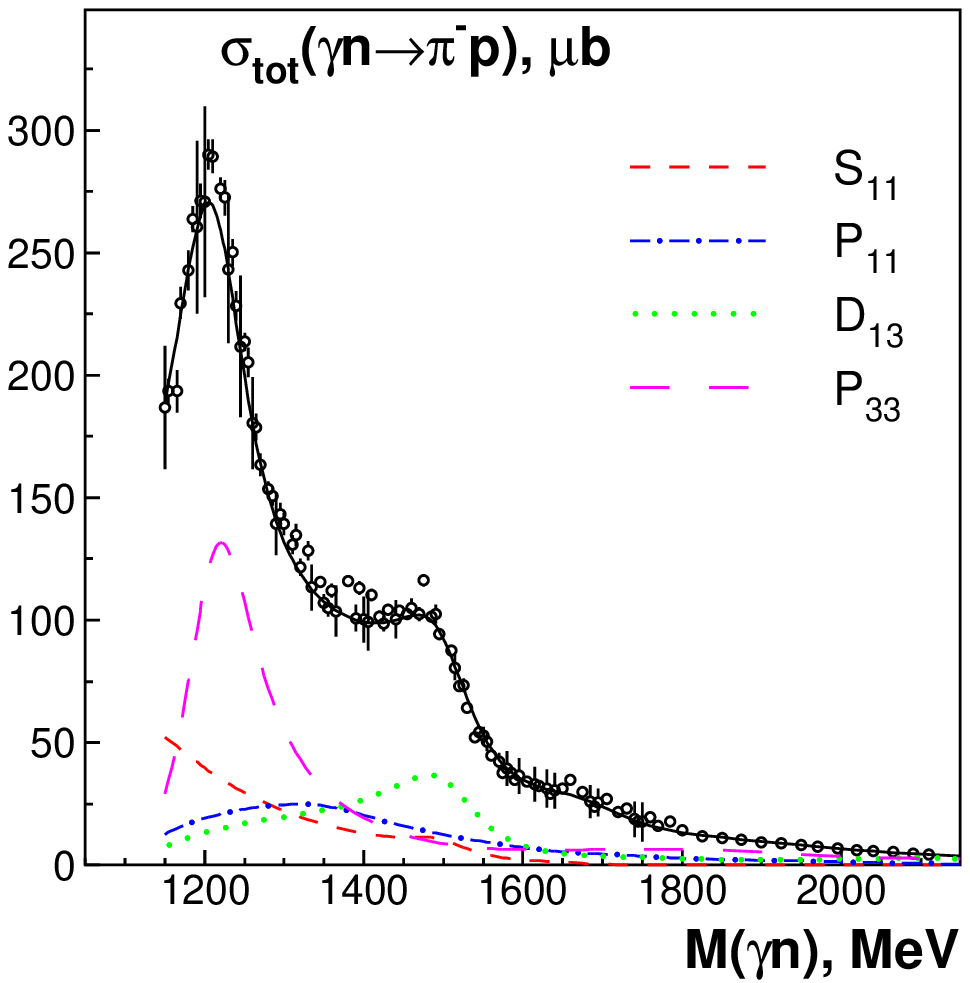,width=0.40\textwidth,height=0.36\textwidth}&
\epsfig{file=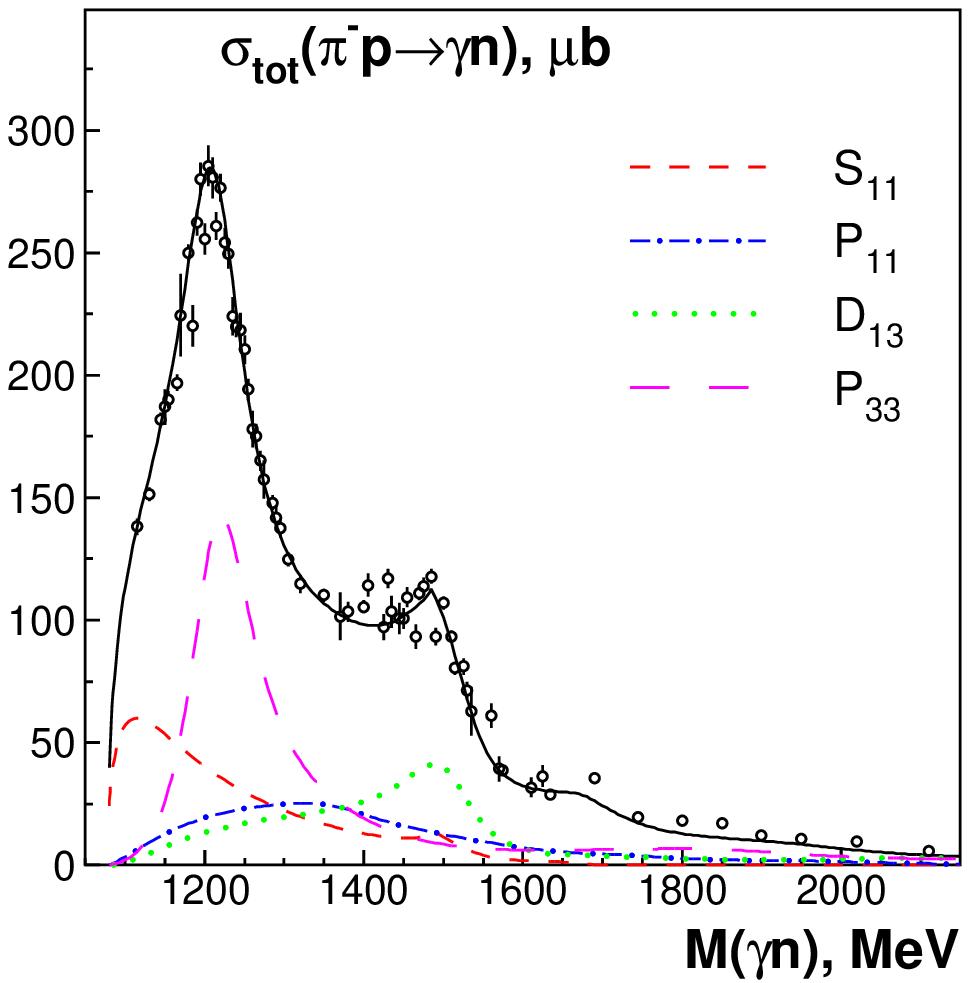,width=0.40\textwidth,height=0.36\textwidth}
\end{tabular}
\end{center}
\caption{Left: Total cross section for $\gamma n\to\pi^- p$. The
Fermi motion is taken into account. Right: Total cross section from
the $\pi^- p\to\gamma n$ reaction, and from data on $\gamma
n\to\pi^- p$ reaction with full kinematics. Corrections for Fermi
motion are not required.
\label{gn_pimp_fm_tot}}
\end{figure*}
\section{Fits to the data}
\begin{figure}[pt]
\centerline{\epsfig{file=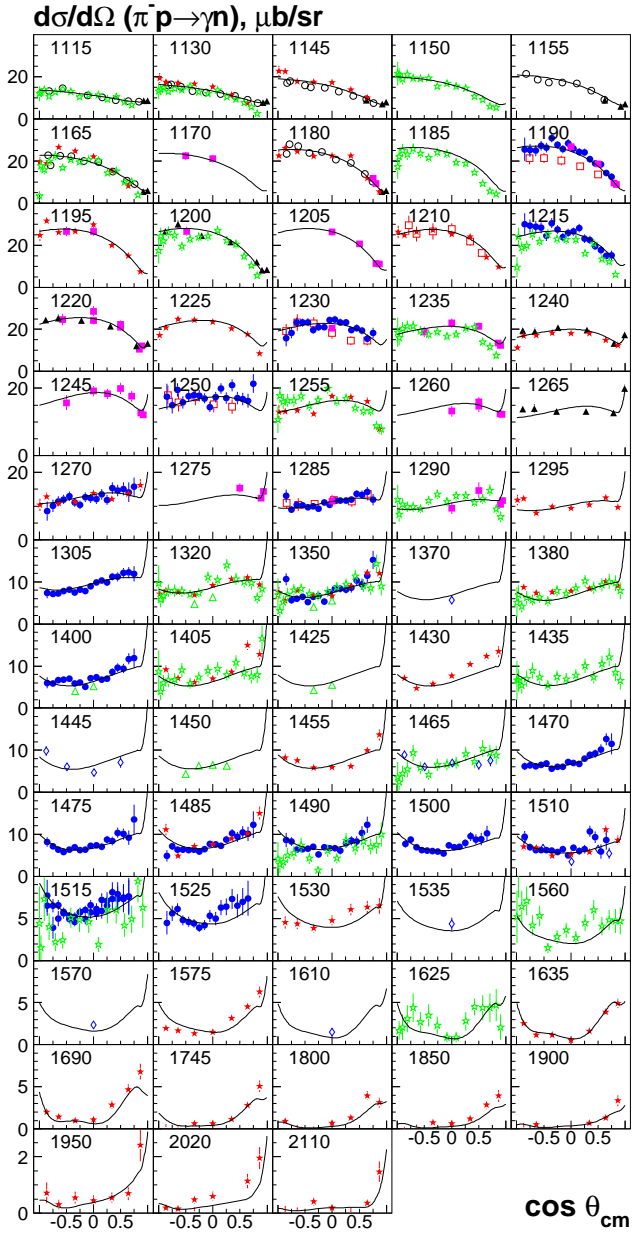,width=0.48\textwidth}}
\caption{Differential cross section for $\pi^- p\to\gamma n$. Data:
black open   circles are from \cite{gn_pion_mp_dcs0}, red open
squares are from \cite{gn_pion_mp_dcs1}, green open triangles are
from \cite{gn_pion_mp_dcs2}, blue  filled circles are from
\cite{gn_pion_mp_dcs3}, pink  filled squares are from
\cite{gn_pion_mp_dcs4}, black filled triangles are from
\cite{gn_pion_mp_dcs5}, red   filled stars   are from
\cite{gn_pion_mp_dcs0_benz}, green open   stars   are from
\cite{gn_pion_mp_dcs0_Rossi}, blue  open diamonds  are from
\cite{gn_pion_mp_dcs6}.
\label{gn_pimp_dcs}}
\end{figure}
As the first step we fitted $\gamma n$ data starting from solutions
BG2011-01 and BG2011-02, respectively. Due to the incompleteness of
the data, several solutions exist which give a similar quality in
the data description. The most significant differences were found in
the $1/2(3/2^+)$ wave where BG2011-02 finds two close-by resonances:
$N(1900)3/2^+$, present in both types of solutions with slightly
different parameters, and $N(1975)3/2^+$, present only in BG2011-02.
But also in the $1/2(5/2^+)$ wave, solutions exist with two or three
poles, the well known $N(1680)5/2^+$ and a second resonance at about
2090\,MeV or, alternatively, $N(1680)5/2^+$ and two resonances at
about 1860\,MeV and 2190\,MeV.

The description of the data, obtained when starting from the
solution BG2011-01 appeared to be systematically worse than that
from the solution BG2011-02. Fig.~\ref{gn_pimp_fm_dcs_1} shows
differential cross sections for $\gamma d\to p \pi^-(p)$ from JLab
\cite{Chen:2012yv} and earlier
measurements \cite{gn_pion_md_dcs1,gn_pion_md_dcs2,%
gn_pion_md_dcs3,gn_pion_md_dcs4,gn_pion_md_dcs5,gn_pion_md_dcs6}.
The data on recoil polarization $P$ and target asymmetry $T$ are
shown in Fig.~\ref{gn_pimp_fm_ppol}, those on the beam asymmetry
$\Sigma$ in Fig.~\ref{gn_pimp_fm_sigm_1}.

\begin{figure}[pt]
\centerline{\epsfig{file=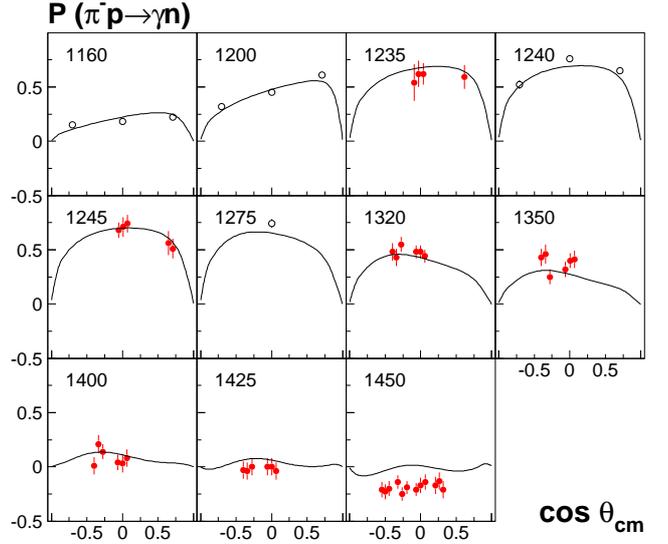,width=0.48\textwidth}}
\caption{Recoil polarization $P$ for $\pi^- p\to n\gamma$. Data:
black open circles are from \cite{gn_pion_mp_ppol}, red filled
circles are from \cite{gn_pion_mp_ppo1}.
\label{gn_pimp_ppol}}
\end{figure}

The data on $\gamma d\to p \pi^- (p)$ are fitted simultaneously with
data on the inverse reaction $\pi^-p\to n\gamma$
\cite{gn_pion_mp_dcs0,gn_pion_mp_dcs1,gn_pion_mp_dcs2,%
gn_pion_mp_dcs3,gn_pion_mp_dcs4,gn_pion_mp_dcs5,gn_pion_mp_dcs6,%
gn_pion_mp_ppol,gn_pion_mp_ppo1} (and data from bubble chambers
\cite{gn_pion_mp_dcs0_benz,gn_pion_mp_dcs0_Rossi}) which are free
from Fermi corrections. The total cross sections for these two
reactions - linked by time reversal invariance - are shown in Fig.
\ref{gn_pimp_fm_tot}. The data points represent the summation over
the differential cross sections; uncovered angular regions are taken
from the fit to the data presented in
Figs.~\ref{gn_pimp_fm_dcs_1}-\ref{gn_pimp_fm_sigm_1}. The data on
$d\sigma /d\Omega$ and on $P$ from the inverse reaction $\pi^-p\to
n\gamma$ are shown in Figs.~\ref{gn_pimp_dcs} and
\ref{gn_pimp_ppol}, respectively.

Resonance contributions to the reaction $\gamma d\to n\pi^0 (p)$ are related to  $\gamma d \to
p\pi^- (p)$ by Clebsch-Gordan coefficients. For the decay of
$\Delta$ resonances, the $p\pi^-$ decay is disfavored by a factor 2
compared to $n\pi^0$, for nucleon resonances, this is reversed. At
the production vertex, the situation is a bit more complicated. The
isospin $3/2$ partial waves are produced with the same couplings in
the $\gamma p$ and $\gamma n$ interaction which imposes particular
relations on the t- and u-exchange amplitudes. In the case of
production of neutral mesons the fitting of the $\gamma n$ and
$\gamma p$ reactions helps to distinguish between (reggeized)
t-channel exchanges with isospin 1 and 0, e.g. $\rho$ and $\omega$
exchanges. In the case of photoproduction of charged pions off
neutrons, the t-channel exchange is fully fixed from the fit to the
reactions with $\gamma p$ in the initial state.

The $\gamma n\to \eta n$ differential cross section shows a peak in
the mass region 1700 MeV which can be described either as an
interference (including cusp-) effect in the $J^P=1/2^-$ partial
wave or as a contribution from a narrow state $N(1680)$ in the
$J^P=1/2^+$ partial wave. In the present analysis every fit was made
with and without contribution from the $N(1680)$ state. All these
solutions were included to define the systematic errors for the
$\gamma n$ couplings.

\begin{figure}
\centerline{\epsfig{file=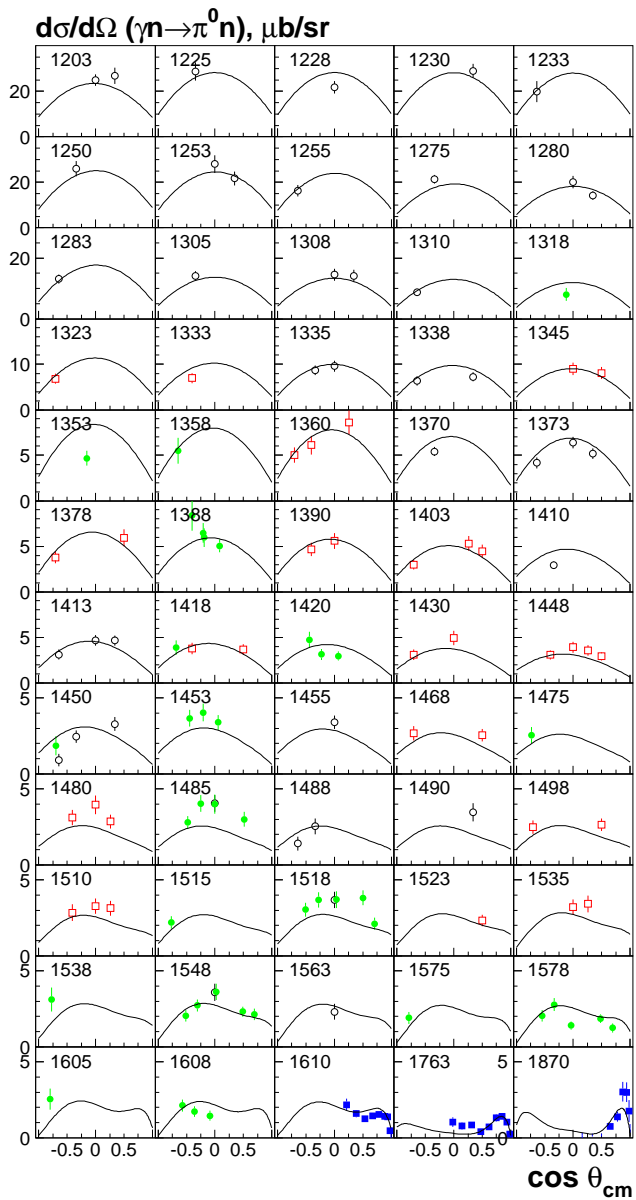,width=0.48\textwidth}}
\caption{Differential cross section for $\gamma n\to\pi^0 n$. Data:
black open circles are from \cite{gn_pion_pd_said}, red open squares
are from \cite{gn_pion_pd_2}, green filled circles are from
\cite{gn_pion_pd_3}, blue  filled squares are from
\cite{gn_pion_pd_4}.
\label{pi0_pd_dcs}}
\end{figure}

\begin{figure}
\centerline{\epsfig{file=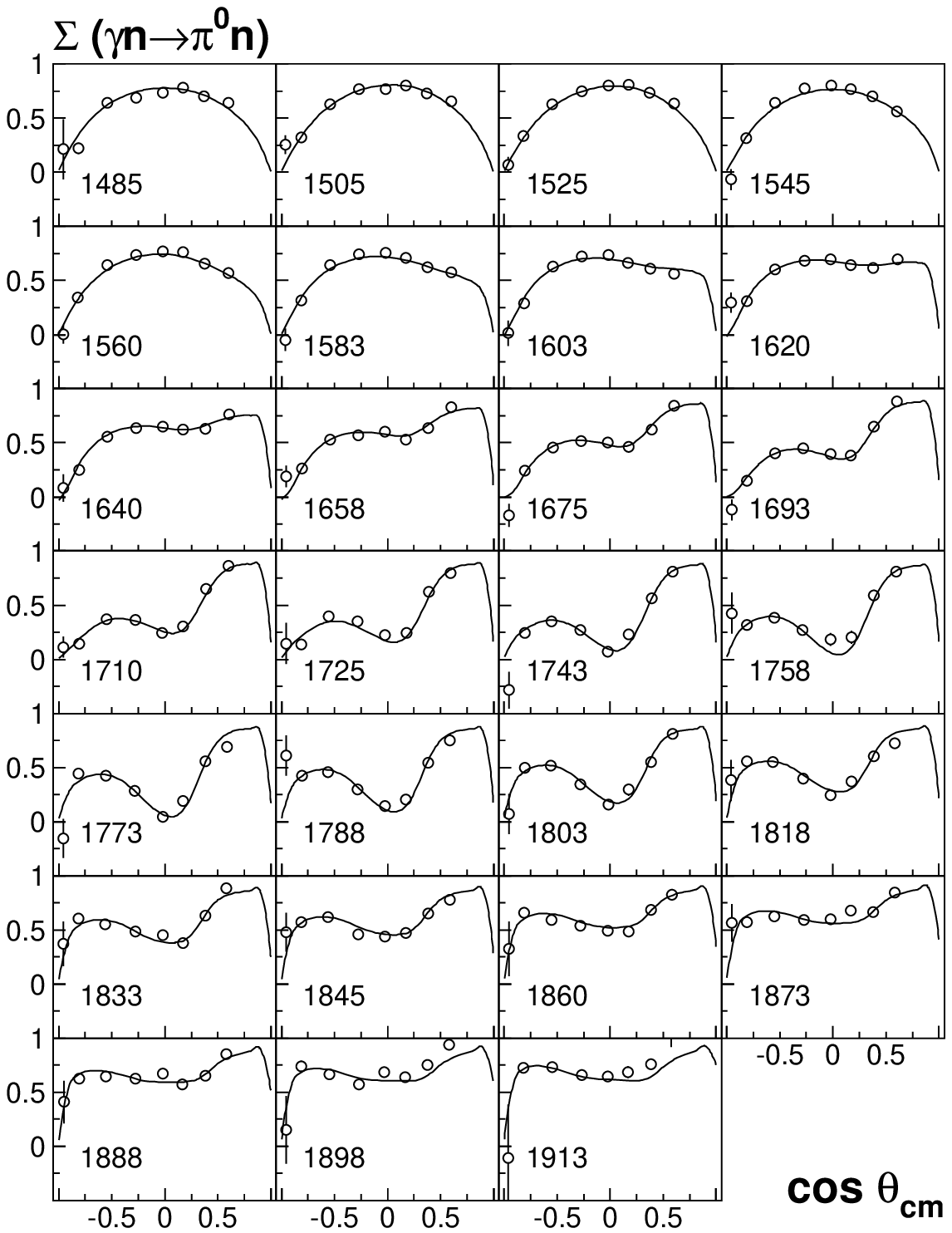,width=0.48\textwidth}}
\caption{$\Sigma$ polarization for $\gamma n\to\pi^0 n$
\cite{gn_pion_pd_gra1}.\vspace{1mm}
\label{pi0_pd_sigma}}
\centerline{\epsfig{file=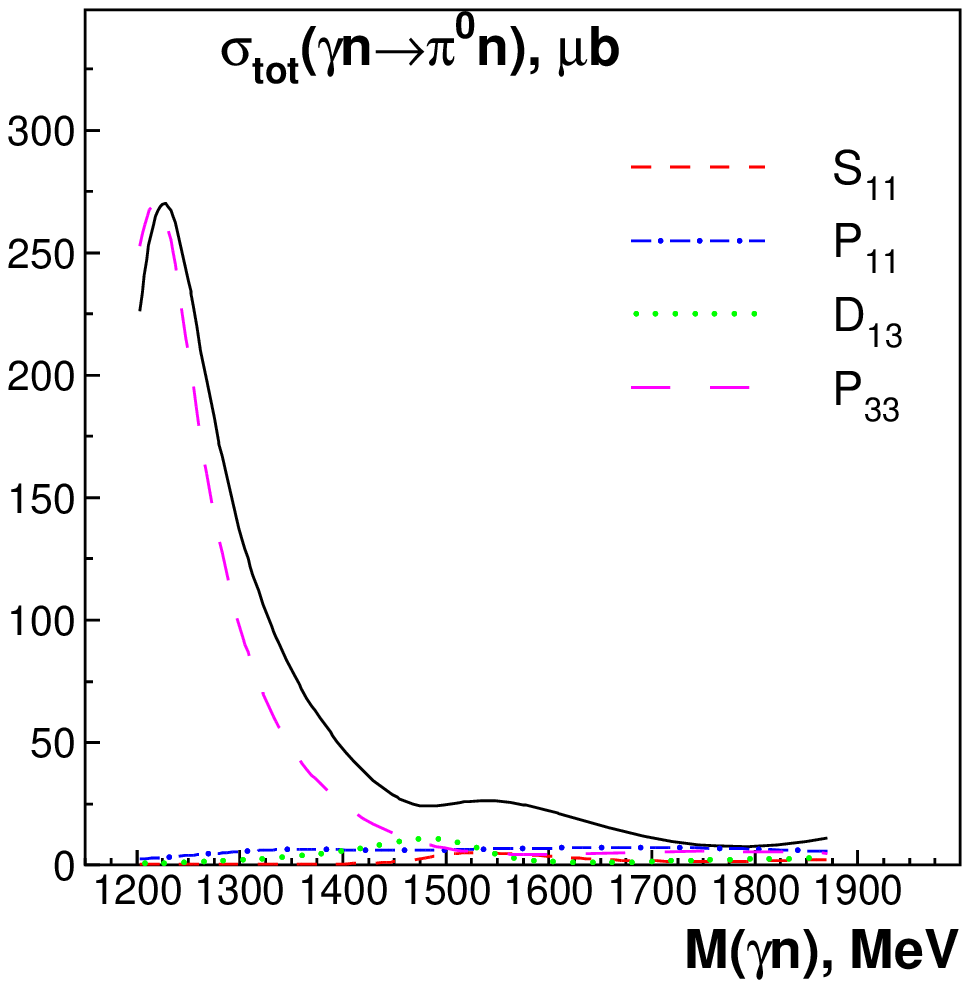,width=0.40\textwidth,height=0.36\textwidth}}
\caption{Total cross section for $\gamma n\to\pi^0 n$.
\label{pi0_pd_tot}}
\end{figure}

\section{The partial wave analysis}
The fits are based on the BnGa partial wave analysis program. The
program is documented in a series of papers \cite{Anisovich:2004zz,%
Anisovich:2006bc,Klempt:2006sa,Anisovich:2007zz}. This paper is
based on our latest solution BG2011-02 \cite{Anisovich:2011fc}.
Here, we mention a few points which are of particular interest for
this work.

\begin{figure}
\centerline{\epsfig{file=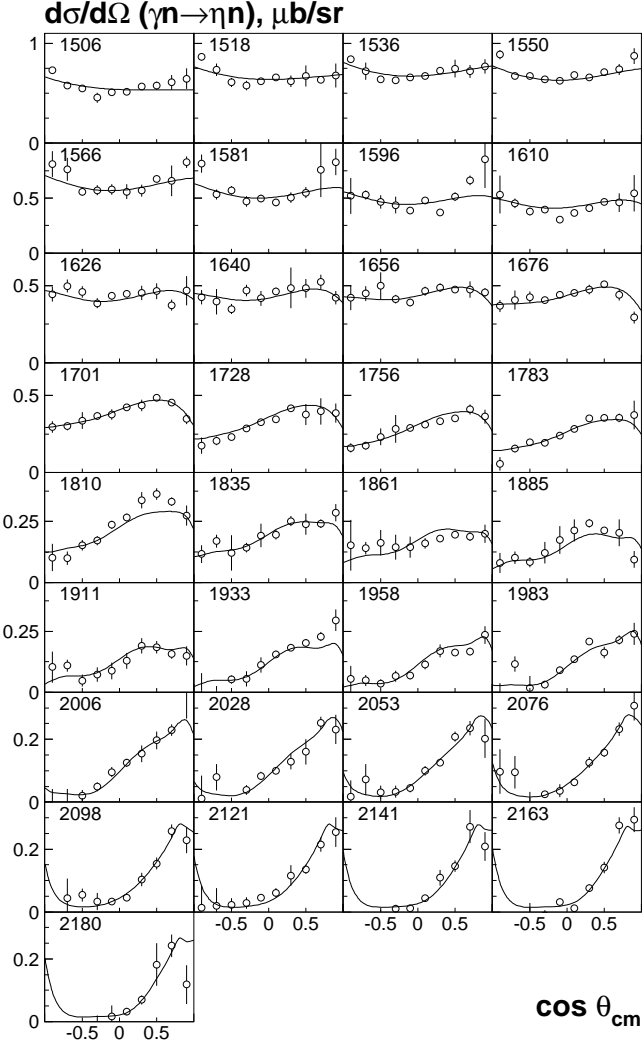,width=0.48\textwidth}}
\caption{Differential cross section for $\gamma n\to\eta n$
\cite{gn_eta0_pd_els2}.
\label{eta_pd_dcs}}
\end{figure}

\begin{figure}
\centerline{\epsfig{file=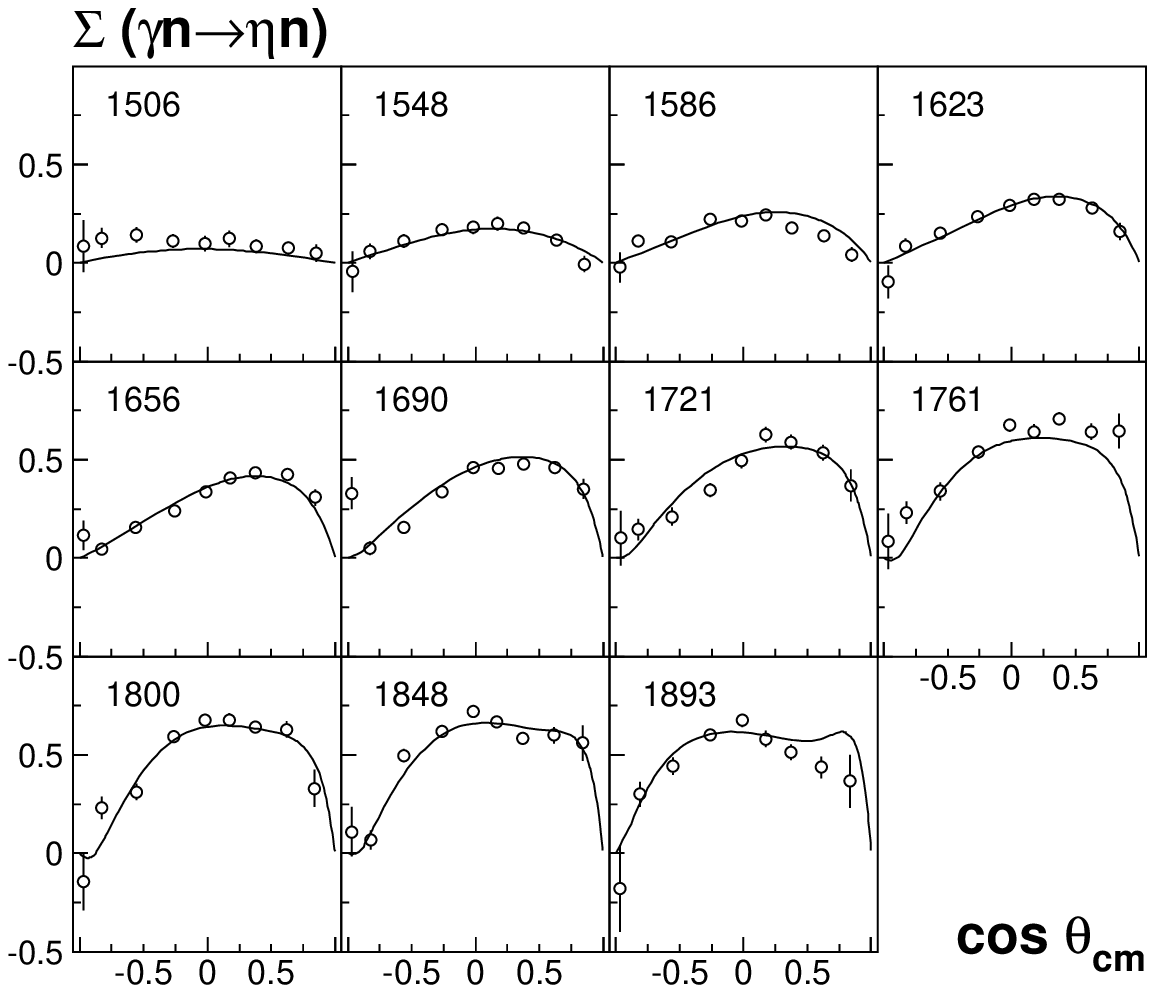,width=0.48\textwidth}}
\caption{$\Sigma$ polarization for $\gamma n\to\eta n$
\cite{gn_eta0_pd_gra1}.\vspace{1mm}
\label{eta_pd_sigma}}
\centerline{\epsfig{file=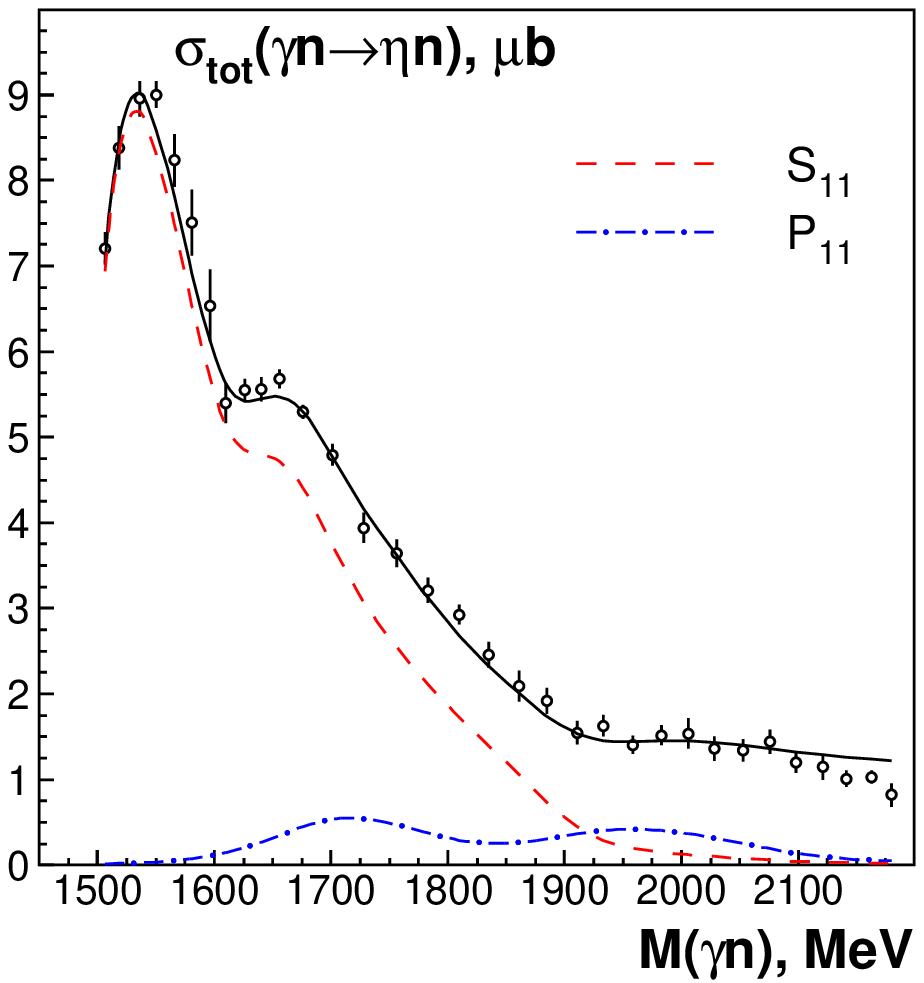,width=0.40\textwidth,height=0.36\textwidth}}
\caption{Total cross section for $\gamma n\to\eta n$.
\label{eta_pd_tot}}
\end{figure}
In the fits, the partial waves (up to $J=7/2$) were described in the
framework of the K-matrix/P-vector approach, while the contribution
of higher partial waves contribute only via Regge $t$ and
$u$-channel exchange amplitudes.  Thus, fixing all parameters
related to the description of the pion and photo-induced reactions,
we obtained a rather good description of the data listed in
Table~\ref{gamma_n_data}. The quality of the description obtained
with the solution BG2011-02 is demonstrated in the fourth column of
Table~\ref{gamma_n_data} and denoted as $\chi_0^2$.

The solution we found was not unique. Starting from different
initial $\gamma n$ couplings for the nucleon resonances we observed
several minima. In one case (like the one shown in
Table~\ref{gamma_n_data}), the differential cross section and beam
asymmetry were well described while the fit to target and recoil
asymmetry was only fair. In other fits the target and recoil
asymmetries were described with a notably better $\chi^2$ while
differential cross section and beam asymmetry had a worse
description providing almost same total $\chi^2$.

A refit of the full data base including pion and photo-induced
reactions off protons with all parameters free improved very notably
the description of the data on photoproduction off neutrons while
the fit to pion and photo-induced reactions using protons did not
deteriorate notably. The improvement of the description of the
$\gamma n$ data can be estimated from the $\chi^2_f$ given in the
last column of Table~\ref{gamma_n_data}.

However, also the full refit of all data did not lead to a unique
solution: there is still a number of acceptable solutions with
rather different $\gamma n$ couplings of the resonances in the third
and fourth resonance region. Probably this is not a big surprise:
most of these states have rather small couplings to the $\pi N$
channel and cannot be reliably defined from the present data. As the
result, in a number of solutions the fit returned for these states
large $\gamma n$ amplitudes with destructively interfering couplings
which provided a slightly better description than smaller and more
realistic amplitudes. When these couplings were fixed to zero, the
result was often only a small deterioration of the description. In
the present analysis, all these solutions were used to estimate the
systematic errors of the neutron helicity couplings. Obviously,
$\gamma n$ couplings of high-mass resonances cannot yet be defined
precisely; future data on photoproduction of open strangeness and of
multi-meson final states and data on double polarization will
certainly improve the situation.

\begin{table*}[pt]
\caption{\label{hel_coup} The $\gamma n$ helicity couplings of
nucleon states (GeV$^{-1/2}10^{-3}$) calculated as residues in the
pole position and corresponding Breit-Wigner couplings. The sign of
a helicity amplitude at the pole position is chosen to have a phase
$<180$\oo . For convenience of the reader we give pole positions
from \cite{Anisovich:2011fc} in last two columns.}
\begin{center}
\renewcommand{\arraystretch}{1.40}
\begin{tabular}{l|ccc|ccc|cc}
\hline\hline ~              & $\pm\,|A_n^{1/2}|$ &Phase    &
$A_{n(BW)}^{1/2}$ & $\pm\,|A_n^{3/2}|$
&Phase & $A_{n(BW)}^{3/2}$ &$M_{\rm pole}$ & $\frac12\Gamma_{\rm pole}$ \\
\hline
$N(1535)1/2^-$ &  -103\er11  &8\er5\oo  & -93\er11 &&&&1501$\pm$4&134$\pm$11  \\
$N(1650)1/2^-$ &  25\er 20   &0\er15\oo &  25\er20 &&&&1647$\pm$4&103$\pm$8   \\
$N(1895)1/2^-$ &  17\er10  & 5\er30\oo   & 13\er6  &&&&1900$\pm$15&90$^{+30}_{-15}$    \\
\hline
$N(1440)1/2^+$ &  35\er12  & 25\er25\oo     &  43\er12 &&&&1370$\pm$4&190$\pm$7        \\
$N(1710)1/2^+$ &  -40\er20  & -30\er25\oo  &  -40\er20 &&&&1687$\pm$17&200$\pm$25    \\
$N(1880)1/2^+$ &  -60\er50  & -30\er40\oo  &  -60\er50 &&&&1860$\pm$35&270$\pm$70      \\
\hline
$N(1520)3/2^-$ &  -49\er8  & -3\er8\oo &  -49\er8    & -114\er12& 1\er3\oo& -113\er12  &1507$\pm$3&111$\pm$5   \\
$N(1700)3/2^-$ &  31\er10   & -50\er30\oo    & 25\er10 &-35\er18 &-30\er30\oo &-32\er18 &1770$\pm$40&420$\pm$180  \\
$N(1875)3/2^-$ & 9\er6  & not def.   &   10\er6    & -19\er15 & not def.& -20\er15  &1860$\pm$23&200$\pm$20  \\
$N(2120)3/2^-$ & 112\er40  &-30\er25\oo   &   110\er45    & 40\er30 &-55\er60\oo &40\er30 &2110$\pm$50&340$\pm$45  \\
\hline
$N(1720)3/2^+$ &  -80\er50  & -20\er30\oo   & -80\er50&  -140\er65    & 5\er30\oo &-140\er65 &1660$\pm$30&450$\pm$100   \\
$N(1900)3/2^+$ & -5\er35  & 30\er30\oo   & 0\er 30 &-60\er40    & 45\er40\oo &-60\er45  &1900$\pm$30&200$^{+100}_{-\ 60}$  \\
\hline
$N(1675)5/2^-$ &  -61\er7  & -10\er5\oo  &  -60\er 7    & -89\er10&-17\er7\oo& -88\er10  &1654$\pm$4&151$\pm$5  \\
$N(2060)5/2^-$ & 27\er12  & -45\er25\oo   &  25\er 11    & -40\er18 &55\er30\oo &-37\er17&2040$\pm$15&390$\pm$25  \\
\hline
$N(1680)5/2^+$ &  33\er6  & -12\er9\oo     &  34\er6    & -44\er9 &8\er10\oo &44\er9  &1676$\pm$6&113$\pm$4       \\
$N(1860)5/2^+$ &-20\er13  & 50\er45\oo   &  21\er13 & 35\er17   &25\er35\oo & 34\er17  &1830$^{+120}_{-\ 60}$&250$^{+150}_{-\ 50}$        \\
$N(2000)5/2^+$ & -17\er12  &-50\er60\oo   &  -18\er12    &-35\er20   &-50\er90\oo &-35\er20   &2030$\pm$110&480$\pm$100   \\
\hline
$N(1990)7/2^+$ &-45\er20  & -50\er35\oo &-45\er20     &  -50\er25    & -45\er40\oo  &-52\er 27  &2030$\pm$65&260$\pm$60     \\
\hline
$N(2190)7/2^-$ &-15\er12  & 50\er40\oo & -15\er13&    -33\er 20      &  25\er20\oo    & -34\er22 &2150$\pm$25&330$\pm$30         \\
 \hline\hline
\end{tabular}
\renewcommand{\arraystretch}{1.0}
\end{center}
\end{table*}

\section{Results}

The helicity couplings calculated from pole residues are given in
Table~\ref{hel_coup} for all fitted resonances together with the
Breit-Wigner couplings reconstructed following the prescription
given in \cite{Anisovich:2011fc}. Due to the multitude of solutions
with acceptable $\chi^2$, the results on neutron helicity amplitudes
vary over some range. The spread of results is used to estimate the
error bands in Table~\ref{hel_coup}. The comparison of these
couplings with results of other analyses is shown in
Table~\ref{bw_coup}. Several of our couplings differ rather notably
from average values given in the review of Particle Data Group
\cite{Beringer:1900zz}. However, our results are often in a good
agreement with the latest analysis of the $\gamma n$ data from the
George Washington University group (SAID) \cite{Chen:2012yv}.

As in our previous analysis of the $\gamma n$ data
\cite{Anisovich:2008wd}, we found a large $\gamma n$ coupling for
$N(1535)1/2^-$. The coupling is larger than the value found in the
SAID solutions SN11 \cite{Workman:2011vb} and GB12
\cite{Chen:2012yv}, and very significantly larger compared to
solution \cite{Arndt:1995ak}. In the latest analysis of the $\gamma
n$ data, the SAID group found an alternative solution GZ12
\cite{Chen:2012yv} which provides a slightly better description of
the data and is fully compatible with our result: see numbers in
parenthesis in Table~\ref{bw_coup}. We remark that the SQTM analysis
\cite{Burkert:2003} predicts a value $A_n^{1/2}= (-90 \pm
5)10^{-3}$GeV$^{-1/2}$ for this state, in excellent agreement with
our results.
\begin{table*}[pt]
\caption{\label{bw_coup} The comparison of our $\gamma n$ helicity
couplings (GeV$^{-1/2}10^{-3}$) for Breit-Wigner resonances with
SAID solution GB12 \cite{Workman:2012jf}, solution SN11
\cite{Workman:2011vb}, MAID solution \cite{Drechsel:2007if},
solution ShMa \cite{Shrestha:2012ep}, PDG average numbers
\cite{Beringer:1900zz}, and SQTM projections \cite{Burkert:2003}. In
the case of GB12 solution \cite{Workman:2012jf} the alternative
solution for the $J^P=1/2^-$ partial wave GZ12 is shown in
parenthesis. }
\begin{center}
\renewcommand{\arraystretch}{1.40}
\begin{tabular}{lcc|lcc}
\hline\hline
~ & $A_{n(BW)}^{1/2}$ & $A_{(BW)}^{3/2}$ & ~ & $A_{(BW)}^{1/2}$ & $A_{(BW)}^{3/2}$ \\
\hline
$N(1535)1/2^-$ &  -93\er11 &  ~     &  $N(1440)1/2^+$ &  43\er12 &  ~     \\
GB12           & -58\er6 (-85\er15)&~ & GB12          &  48\er4  &  ~     \\
SN11           & -60\er3  &  ~      &   SN11          &  45\er15 &  ~     \\
MAID           &  -51     &         &   MAID          &  54      & ~    \\
ShMa           &  -49\er3 &         &   ShMa          &  40\er5  & ~    \\
PDG12          & -46\er27 &  ~      &   PDG12         &  40\er10 &  ~     \\
SQT03 & -90\er 6 & ~     & ~ & ~ & ~ \\
\hline
$N(1650)1/2^-$ &  25\er20 &  ~     &  $N(1710)1/2^+$ & -40\er20  &  ~     \\
GB12           & -40\er10 (?\er?)&~ &   GB12       & ~      &  ~     \\
SN11           & -26\er8  &  ~      &   SN11         & ~      &  ~     \\
MAID           &  9       &   ~     &   MAID         & ~       & ~ \\
ShMa           &  11\er2  &   ~     &   ShMa         & 17\er3  & ~ \\
PDG12          & -15\er21 &  ~      &   PDG12        & 2\er14 &  ~     \\
SQT03 &  -31\er3 & ~     & ~     & ~     & ~ \\
\hline\hline
$N(1520)3/2^-$ & -49\er8 & -113\er12      & $N(1720)3/2^+$ &  -80\er50    & -140\er65   \\
GB12           & -46\er6 &-115\er 5       &  GB12               &  ambiguous &ambiguous\\
SN11           & -47\er2 &-125\er 2       &  SN11               &  -21\er4 &-38\er 7\\
MAID           & -77     & -154           &  MAID               &   -3     & -31 \\
ShMa           & -38\er3 & -101\er4       &  ShMa               &   -2\er1 & -1\er2 \\
PDG12          & -59\er9 &-139\er 11      &  PDG12              &  4\er15 &-10\er 20\\
SQT03 & -44\er7  &   -140\er 5 & ~  & ~ & ~ \\
\hline \hline
$N(1675)5/2^-$ & -60\er7 & -88\er10   &$N(1680)5/2^+$&  34\er6 & -44\er9   \\
GB12           & -58\er2 &-80\er 5   &GB12           &  26\er4  &-29\er 2  \\
SN11           & -42\er2 &-60\er 2   &SN11           &  50\er4  &-47\er 2  \\
MAID           &  -62    & -84       &MAID           &  28      & -38     \\
ShMa           &  -40\er4& -68\er4   &ShMa           &  29\er2  & -59\er2  \\
PDG12          & -43\er12 &-58\er 13 &PDG12          &  29\er10 &-33\er 9\\
SQT03 & -38\er 3 & -53\er 8    & ~ & ~ & ~\\
 \hline\hline
\end{tabular}
\renewcommand{\arraystretch}{1.0}
\end{center}
\end{table*}

For the second $1/2^-$ state $N(1650)$, our fit optimized for a
positive helicity coupling while most other analyses - our own
previous result \cite{Anisovich:2008wd} and both recent SAID results
\cite{Chen:2012yv,Workman:2011vb}  - report negative values. In the
present fits, this coupling optimized always at a positive value.
This result seems to be driven by the different interference effect
in $\eta$ photoproduction from $\gamma p$ and $\gamma n$ initial
states: in $\gamma n\to n\eta$, a peak structure at 1680\,MeV is
observed in the $N\eta$ invariant mass which is absent in $\gamma
p\to p\eta$. However, even in solutions where this peak structure
was described as a contribution from $N(1680)1/2^+$, and some
interference in the $1/2^-$ partial wave, the $N(1650)1/2^-$
helicity coupling became smaller but remained positive. To
investigate systematically our result we started fits from our
solutions reported in \cite{Anisovich:2008wd} and fitted with free
parameters or fixed to zero the couplings of high mass states which
were not included in the fit at that time. However, in all cases the
present fits optimize for a positive coupling. This result was
reproduced by both solutions BG2011-01 and BG2011-02.

In our fits, the opposite signs of the $N(1650)1/2^-$ and
$N(1535)1/2^-$ helicity amplitudes are required when the dip-bump
structure in the $\gamma n\to n\eta$ total cross section in
Fig.~\ref{eta_pd_tot} is assigned to the $J^P=1/2^-$ wave. If the
two helicity amplitudes are forced to have the same sign, the
peak-bump structure is described by $N(1710)1/2^+\to n\eta$ decays.
The description is improved when a narrow $N(1685)$ resonance
\cite{Kuznetsov:2010as,Jaegle:2011sw} is admitted in the fit in
addition to $N(1710)$, yet the overall $\chi^2$ remains
significantly worser than in the case where opposite signs of the
$N(1650)1/2^-$ and $N(1535)1/2^-$ helicity amplitudes are admitted.
Opposite signs for $N(1650)1/2^-$ and $N(1535)1/2^-$ helicity
amplitudes are also found in the recent analysis by Shrestha and
Manley \cite{Shrestha:2012ep}. The MAID solution
\cite{Drechsel:2007if} is compatible with our value as well. The
SAID group did not report the exact number for the GZ12 solution:
only magnitude of the coupling was mentioned ($\approx
20$\,GeV$^{-1/2}10^{-3}$ \cite{Workman:2012jf}).

We mention that our old (now superseded) result was derived from an
insufficient data base. A fit with a free helicity coupling for
$N(1650)$ did not converge properly. Hence the direct coupling was
set to zero. Through rescattering, a finite (negative) value for the
$N(1650)$ helicity coupling emerged. Now, the data base has much
improved, and we do no longer quote helicity amplitudes which led to
a bad convergency property of the fit (except for error studies).

The $\gamma n$ coupling of the  Roper resonance $N(1440)1/2^+$ was
found to be in a good agreement with results of other analyses. The
fit also optimized at a rather large $\gamma n$ couplings for the
$N(1710)1/2^+$ and $N(1880)1/2^+$ states, the latter with very large
uncertainties. However, there is a strong interference between these
couplings: both states contribute very little to $\pi$ and $\eta$
photoproduction and therefore their couplings cannot be defined
reliably from the analysis of the present data set. If the primary
photo production coupling of both higher $1/2^+$ states are fixed to
zero, the description does not change significantly. Even in this
fit we found - via rescattering - rather large $\gamma n$ couplings
as pole residues. All these solutions were included to define
systematic errors for couplings given in Table~\ref{hel_coup}.

Both $N(1520)3/2^-$ $\gamma n$ couplings were found to be notably
smaller than the average PDG values while our present result is
fully compatible with the latest fit of the SAID group; their values
are thus confirmed by our analysis. We found rather small helicity
couplings for $N(1875)3/2^-$: the fit does not demand a contribution
from this resonances. Also the neutron helicity amplitudes for
$N(1700)3/2^-$ are not well fixed by the data. The $A_n^{1/2}$
coupling of the $N(2120)$ $3/2^-$ state was found to be rather large
and stable in the fit. However in this energy range there are only
good data in the photoproduction of charged pion \cite{Chen:2012yv}.
Therefore a presence of a $\Delta$ state with $J^P=3/2^-$ can change
the situation significantly.

Our result for the helicity amplitudes of $N(1675)5/2^-$ is also
rather different from the average PDG values but again coincides
remarkably well with the latest SAID analysis GB12 and with MAID.
The SQTM projections of -38 $\pm$ 5 and -53 $\pm$ 8 agree with PDG
averages as well as with SN11 but are at
variance with our result and GB12. We also believe that we have
obtained reliable numbers for $N(2060)5/2^-$: in our solutions this
state contributes to both $\eta$ and $\pi$ photoproduction
reactions.

Our result for the $A_n^{3/2}$ helicity coupling of $N(1680)5/2^+$
deviates by more than one standard deviation from the SAID fit GB12,
but is in a good agreement with the SN11 solution. Probably, there
is some freedom in this region which should be fixed by future
double polarization data. We found rather small couplings for the
higher $5/2^+$ states. The large systematic errors here are not
surprising: these couplings are difficult to define without
including other final states like $\Lambda K$ or $N\pi\pi$ and
double polarization data.

The fits based on the solution BG2011-02 produced a notably better
description of the data than fits based on the solution BG2011-01.
Hence we provide the $\gamma n$ couplings for $N(1990)7/2^+$
obtained with the solution based on BG2011-02. Excluding these
couplings from the fit leads to a deterioration of the description
in the high mass region, hence we consider these results as
reliable.

We have found a rather small improvement from the $\gamma n$
couplings of the $N(2190)7/2^-$ state, and the couplings were almost
compatible with zero.

The size of the photocouplings of the $N(1720)3/2^+$ is a highly
controversial topic. Already our analysis of $\gamma p$ interactions
showed a large discrepancy between our result and the mean PDG value
\cite{Beringer:1900zz}. The discrepancy was partly resolved by the
latest SAID analysis \cite{Workman:2011vb} where the $A_p^{1/2}$
coupling was found to be compatible with our result. But our
$A_p^{3/2}$ value is still more than three times larger than the
SAID result. In our present analysis we found $\gamma n$ couplings
which are also dramatically larger than those found in the SAID
analysis SN11 \cite{Workman:2011vb}. In the analysis GB12-GZ12 the
couplings for $N(1720)3/2^+$ are not given: the fit found a number
of solutions with rather different contributions from this state.

One of the problems may be related to the complicated structure in
this partial wave; a second problem is due to the possibility to
assign intensity in the 1700\,MeV region to $N(1720)3/2^+$, to
$N(1710)1/2^+$ or to $N(1700)3/2^-$/ $\Delta(1700)3/2^-$. A third
source for the differences in the results for the $N(1720)3/2^+$
helicity amplitudes could be the difference of the methods used to
extract resonance parameters. This state is rather broad and highly
inelastic: the elasticity is about 10\% only. The elastic pole
residue has a very large phase: it varies between $-90^\circ$ and
$-140^\circ$. The phase reflects a large contribution from
non-resonant terms which make it difficult to treat the state as
Breit-Wigner resonance. The amplitude pole corresponding to this
state is located in the 1660-1690\,MeV region, and the dominant
decay into the $2\pi p$ final state is complicated due to the
opening of the $\pi N(1520)3/2^-$ channel (which was found to be
rather strong in a number of our solutions). Further, this resonance
is close to $\Delta(1600)3/2^+$, a resonance with parameters which
are also not firmly defined. The presence of $N(1900)3/2^+$ and
$N(1875)3/2^-$ makes the picture even more complicated. These
uncertainties contribute to the large errors for the $N(1720)3/2^+$
helicity amplitudes.

\section{Conclusion}

Starting from solution BG2011-02, we obtained a rather good
description of the $\pi$ and $\eta$ photoproduction data off
neutrons which was improved further by allowing the fit to adjust
slightly also masses, widths, coupling constants, and background
terms in an overall fit which includes also data on photoproduction
off protons and pion-induced reactions. The results for the helicity
couplings for photoproduction off neutrons were presented and
compared to results from other analyses.

The couplings of several 4-star resonances differ notably from
average PDG values but are mostly in very good agreement with the
latest SAID analysis GB12(GZ12) \cite{Chen:2012yv}. For the
$N(1535)1/2^-$ state our coupling is fully compatible with one of
the SAID solutions (GZ12). The $N(1650)$ $1/2^-$ helicity amplitude
differs from our previous result \cite{Anisovich:2008wd} and from
results of the two SAID solutions SN11 and GZ12. The $N(1720)3/2^+$
helicity amplitudes are controversial (as they are in the case of
proton helicity amplitudes). More data are required to resolve these
ambiguities. We also have defined the $\gamma n$ helicity couplings
for radial excitations in the third and fourth resonance region.
However in most cases these numbers should be taken with care and
may rather be considered as initial values for a future analysis
when new data with other final states will be included into the fit.

\subsection*{Acknowledgements}
We would like to thank the members of SFB/TR16 for continuous
encouragement. We acknowledge support from the Deutsche
Forschungsgemeinschaft (DFG) within the SFB/TR16. V. Burkert
acknowledges support from the US Department of Energy under contract
DE-AC05-06 OR23177.

\end{document}